\documentclass[a4paper,11pt]{article}
\usepackage{pos}

\title{Probing parton distributions in ep/eA and ultra-peripheral collisions}
\author[a]{Spencer R. Klein}

\affiliation[a]{Lawrence Berkeley National Laboratory,\\
  Berkeley CA 94720 USA}

\emailAdd{srklein@lbl.gov}

\abstract{Real or virtual photons are excellent probes of nuclear structure, with a strong sensitivity to gluon distributions.  Photonic reactions can be studied using ultra-peripheral collisions or at an electron-ion collider.  Final states like dijets or open charm production are directly sensitive to the gluon distributions in nuclei.    Exclusive reactions, like exclusive vector meson production or deeply virtual Compton scattering (DVCS) go further, requiring at least two gluons.  In the Good-Walker paradigm, coherent exclusive photoproduction is sensitive to the average nuclear configuration (including gluonic hot spots), and the Fourier transform of the differential cross-section $d\sigma/dt$ gives the transverse distribution of partonic targets in a nucleus.  The incoherent photoproduction cross-section is sensitive to partonic fluctuations, including gluonic hot spots.   Some reactions, such as dijet production, involve multiple momentum scales, and thus may be able to probe the Wigner distribution of nuclear targets.  Finally, incoherent photoproduction is sensitive to partonic fluctuations; an analysis of $J/\psi$ photoproduction on proton targets found that the data clearly preferred a fluctuating lumpy proton.
}

\FullConference{25th International Spin Physics Symposium (SPIN 2023)\\
 24-29 September 2023\\
 Durham, NC, USA\\}

\begin{document}
\maketitle

\section{Introduction}

Photoproduction has long been used to study nuclear structure \cite{Bauer:1977iq,Alvensleben:1970uv}.  High-energy photons are sensitive to the partonic structure of matter at low Bjorken$-x$ \cite{Klein:2019qfb}.    Although most early photoproduction experiments ran at fixed-target accelerators, interest has shifted toward higher energies which require colliders, either ultra-peripheral collisions (UPCs) at hadron colliders \cite{Baltz:2007kq,Bertulani:2005ru,Contreras:2015dqa,Klein:2020fmr}, or $ep$/$e$A collisions, at HERA or the future U. S. electron-ion collider \cite{AbdulKhalek:2021gbh}.

UPCs are the energy frontier for photon physics.  Heavy nuclei carry strong electromagnetic fields which, per Weizs\"acker-Williams approach can be treated as a flux of nearly-real photons.  Because the photon flux scales with the square of the nuclear charge ($Z$), UPCs are most prominent in collisions involving heavy ions.  The maximum photon energy is set by the size of the nucleus (with radius $R_A$) and the ion's Lorentz boost, $\gamma$.  In the lab frame, the maximum photon energy is $\gamma\hbar c/R_A$.  For lead nuclei at LHC energies, this is about 100 GeV, equivalent to  500 TeV in the rest frame of the target nucleus, giving photon-nucleus center of mass energies up to 700 GeV per nucleon.  For photons emitted by protons, the energies are about 5 times higher, due to the larger $\gamma$ and smaller $R_A$.  These energies allow the LHC to probe partons down to Bjorken$-x\approx 10^{-6}$ at moderate $Q^2$.   

With heavy nuclei,  $Z\alpha$ is large, so it is possible, even frequent, for multiple photons to be exchanged.  This is useful for selecting different impact-parameter ($b$) distributions, and, with that, different photon energy spectra.  However, it also complicates the selection of 'exclusive' interactions.  Another complication arises because the photon direction is ambiguous: either nucleus can emit a photon \cite{Klein:1999qj}.  For exclusive production of a vector meson with mass $M_V$ at rapidity $y$, the two photon energy ($k$) solutions and associated Bjorken$-x$ values are
\begin{equation}
k = \frac{M_V}{2}\exp{(\pm y)}, \hskip .25 in xm_p = \frac{M_V}{2}\exp{(\mp y)}
\label{eq:kandx}
\end{equation}
where $m_p$ is the proton mass.  The two-fold ambiguity can be resolved by analyzing multiple considering classes of events, containing different requirements on additional photon exchange ({\i. e.} with different impact parameter distributions and photon spectra), creating a set  of linear equations that can be solved to find the cross-sections for the two photon directions.    For photoproduction, the $Q^2$ scale is set by the invariant mass of the final state.  For charmonium and bottomonium, this hard scale allows for comparisons with perturbative QCD.  

Electron-ion collisions offer several advantages over UPCs. Photons can have a wide range of virtualty.  By tagging the outgoing electron, the photon 4-momentum can be determined, irrespective of the final hadronic state.  High-energy electron-proton collisions were first studied at the HERA $ep$ collider.  These studies will be continued by the U. S. electron-ion collider (EIC).  The EIC will accelerate nuclei from protons to uranium, at a very high luminosity.  Protons and light ion beams can be polarized, as can the electron beam.  The detector(s) will be optimized for electron-ion collisions, with nearly $4\pi$ angular coverage.  However, the maximum photon energy is lower than is available with UPCs at the LHC.

\section{Experimental Probes}

Photons can produce a wide variety of final states.  Many of these reactions can be described using a dipole model, in which a photon fluctuates to a quark-antiquark dipole, which then interacts hadronically with the target.   

One of the simplest reactions, photon-gluon fusion, can lead to dijets or to open charm.   Since only a single gluon is involved, the final state cannot be color neutral; additional particles must be created by the color string between the dipole and the hard reaction products.  These processes are theoretically simpler than exclusive reactions, but are more complex experimentally.  The dijet or charm-quark pair masses fill the role of the vector meson mass in Eq. \ref{eq:kandx}.  For dijets, the requirement that the jets be energetic enough to be well reconstructed limits the reach in $x$.  The ATLAS collaboration has studied UPC photoproduction of jets in lead-lead collisions \cite{bgilbert}. With the kinematic cuts, the analysis was sensitive to the range $10^{-2} < x < 1$ and 1600 GeV$^2 < Q^2 < 40,000$ GeV$^2$.  The data were consistent with the nCTEQ nuclear parton distribution functions.   Similar analyses are possible for open charm, offering sensitivity to somewhat lower $x$ and $Q^2$ ranges.  Open charm analyses might be able to probe gluons with Bjorken$-x$ down to $\approx 2\times10^{-6}$, at  $Q^2\approx {\rm few}\ m_c^2$.

\section{Exclusive vector meson production}

In exclusive photoproduction a photon fluctuates to a $\overline q q$ dipole, which then scatters elastically but hadronically from a target nucleus, often emerging as a vector meson.  This elastic scattering can be described as occurring via Pomeron exchange.   In lowest order pQCD, the Pomeron consists of two gluons.  Because this involves elastic scattering, the Pomeron has the same quantum numbers as the vacuum; the vector meson retains the quantum numbers of the  photon, including helicity.   Due to the two gluon exchange, this reaction can be described in terms of a Generalized Parton Distribution (GPD).  However, it can be related to single-gluon distribution using a Shuvayev transformation \cite{Jones:2013pga}.  Generally, one gluon carries most of the momentum, and the other is much softer.

Recent next-to-leading order QCD calculations of $J/\psi$ production have found a rather different picture than the LO approach \cite{Eskola:2022vpi}.  The quark contribution is significant, partly because the amplitudes from NLO gluon diagrams have a different sign than the LO contributors.  These data can still be used in NLO parton distribution fits, but the fits must consider both the quark and gluon contributions.  Also, there is an unexpectedly large scale uncertainty.  For now, this can be controlled by comparing the cross-sections on proton and heavy-ion targets, to find the nuclear suppression.    These issues 
%of the large quark contribution and the scale uncertainty also 
also remain in fits to find GPDs using this data.

\begin{figure}[t]
\includegraphics[width=.48\textwidth]{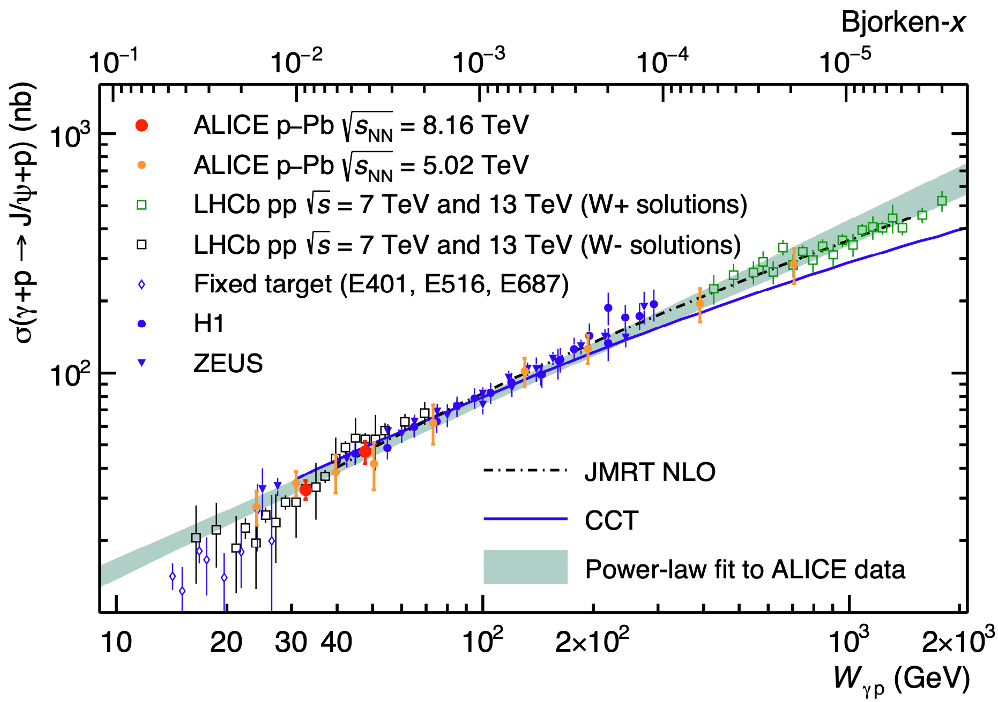}
\includegraphics[width=.48\textwidth]{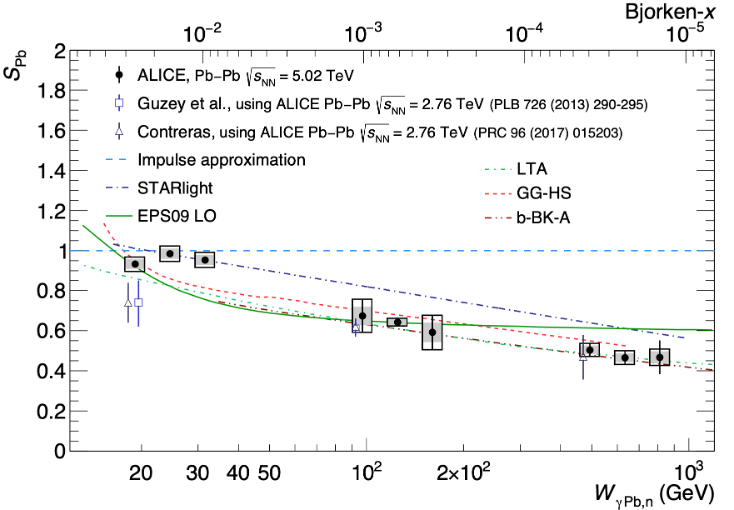}
\caption{(left) The cross-section for $J/\psi$ photoproduction on proton targets. From Ref. \cite{ALICE:2023mfc}. (right) $S_{\rm Pb}$, the $J/\psi$ photoproduction cross-section relative to a proton reference.  As the center-of-mass energy increases,  $S_{\rm Pb}$ drops.  From Ref. \cite{ALICE:2023jgu}.}
\label{fig:gammap}
\end{figure}

Measurements of $J/\psi$ photoproduction on proton targets have been made at the fixed target experiments, HERA, and the LHC, with the latter  using $p{\rm Pb}$ and $pp$ collisions.  In $pA$ collisions, the photon is predominantly emitted by the ion, while for $pp$ collisions, bootstrapping (using other measurements at lower energy) has been used to solve the two-fold ambiguity (Eq. \ref{eq:kandx}).   The cross-section largely follows a power law, with
\begin{equation}
\sigma(W_{\gamma p}) \propto W_{\gamma p}^{0.70 \pm 0.04}
\label{eq:gammap}
\end{equation}
as is shown in Fig. \ref{fig:gammap} (left) \cite{ALICE:2023mfc} .  Except near threshold, the cross-section is close to a power law up to $W_{\gamma p}=2$ TeV, corresponding to Bjorken $x$ around $2\times10^{-6}$ (at $Q^2\approx 2.25$ GeV$^2$).   At lowest order, this is expected if the gluon distributions follow a power law.  However, at NLO, the picture must be more complex.

For lead targets, the cross section $J/\psi$ photoproduction is smaller than for 208 independent nucleons (the impulse approximation prediction)\cite{LHCb:2022ahs,CMS:2023snh,ALICE:2023jgu} .  The suppression factor ($S_{\rm Pb}$) relative to the impulse approximation quantifies the nuclear shadowing, due to changes in the parton distributions.  As Fig. \ref{fig:gammap} (right) shows, $S_{\rm Pb}$ is significantly below 1.    It is also below the predictions of the STARlight Monte Carlo, which uses a Glauber calculation to account for the fact that the nucleons are clumped together into a single nucleus, so photons hitting the center of the nucleus may interact with multiple nucleons, indicating that the modifications go beyond geometric effects.   The data are in agreement with dipole model calculations.  There are many different dipole model calculations, using different models of partons in nuclei, with different gluon shadowing and saturation, and somewhat different predictions.  Unfortunately,  the limited precision of the data and the similarity of different models preclude more specific conclusions.  To make progress, it will be necessary to compare the theories with multiple observables.   The $\psi '$ is heaver than the $J/\psi$, but with a different wave function, so its cross section provides an additional discriminant.  The $\Upsilon$ states should have further discriminating power.   Measurements of $\sigma/dt$ are also sensitive to saturation, which changes the effective shape of the nucleus \cite{Guzey:2016qwo}.

In the Good-Walker picture, coherent and incoherent production of vector mesons are respectively sensitive to the average nuclear configuration and their fluctuations \cite{Good:1960ba,Miettinen:1978jb}.  The coherent cross-section is obtained by summing the amplitudes for the target to remain in its ground state and then squaring, while the total cross-section is obtained by squaring all of the amplitudes and then adding.  The incoherent cross-section is the difference - the square of sums minus the sum of squares, giving a measure of fluctuations.  This approach has been use to study $J/\psi$ photoproduction in $ep$ collisions at HERA, where a model based on a lumpy proton fit the data much better than a smooth proton \cite{Mantysaari:2016ykx}.   However, there are two issues with this Good-Walker approach \cite{Klein:2023zlf}.  

First, Good-Walker associates coherent photoproduction only with interactions where the target remains intact.  However, coherent photoproduction has been observed in two contexts when the target dissociates:  photoproduction accompanied by mutual Coulomb excitation \cite{STAR:2002caw,ALICE:2020ugp}, and photoproduction in peripheral heavy ion collisions \cite{STAR:2019yox,ALICE:2022zso}.    

Second, nuclear excitation is endothermic, so requires enough energy transfer to the target.  Lead and gold are very different in this respect.  The lowest excited state in lead is at an energy of 2.6 MeV, while for gold, the lowest lying state is at 77 keV.  So, when the energy transfer is small, we expect substantial differences in excitation.  In contrast, in the Good-Walker based calculations, lead and gold are similar, with similar nuclear shapes, and similar parton distributions, so should have similar $d\sigma_{\rm incoherent}/dt$.  

These issues can be at least partially avoided in a semi-classical approach \cite{Klein:2023zlf}, where one sums the amplitudes for reactions that are indistinguishable from each other, with a propagator to account for different locations for different target nucleons.  This leads to a similar picture for coherent photoproduction, but a rather different take on incoherent production.  In this model, a completely static target will still have incoherent interactions, although a black disk will not.  

\section{GPDs, and the transverse distribution of targets}

Although vector meson photoproduction is naturally described using GPDs, the data does not permit direct extractions, because $\Delta$, the momentum difference between the two gluons is not known ({\it i. e.} we cannot separate $x_1$ and $x_2$), so there is not direct access to GPDs.   So, recent studies have focused on a related subject, the spatial dependence of the gluon distributions.    The transverse distribution of targets, $F(b)$ is given by the Fourier transform of $d\sigma/dt$:
\begin{equation}
F(b) \propto \int_0^\infty  p_T dp_T J_0(bp_T)   \sqrt{\frac{d\sigma}{dt}}.
\label{eq:fourier}
\end{equation}

Several caveats apply here.  

First, the square root converts cross-section to amplitude.  So, it is necessary to flip the sign of $\sqrt{d\sigma/dt}$ when crossing each diffractive minimum to account for the phase evolution with $t$.  This is theoretically straightforward, but it is not always easy to pinpoint the minima accurately, especially in the presence of final state radiation.

Second, the $p_T$ integral runs from 0 to infinity, but the experimental data is limited to a finite range.  The limited range can introduce windowing artifacts, since the finite integral gives a Fourier transform that is the convolution of $d\sigma/dt$ with a box function.  

\begin{figure}[t]
\includegraphics[width=.38\textwidth]{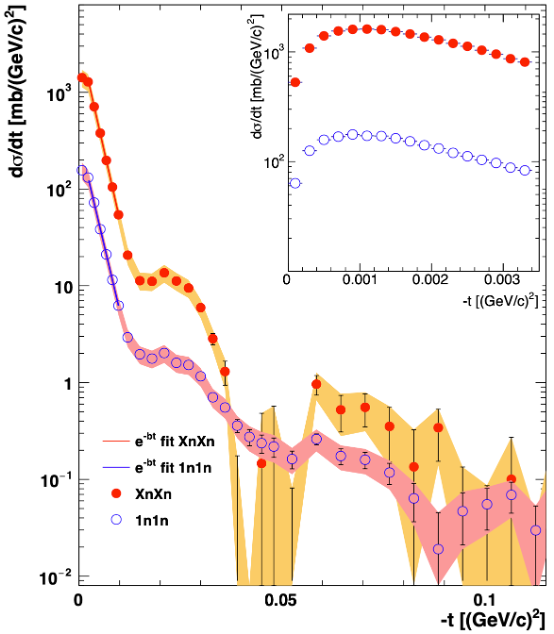}
\hskip .1 in
\includegraphics[width=.48\textwidth]{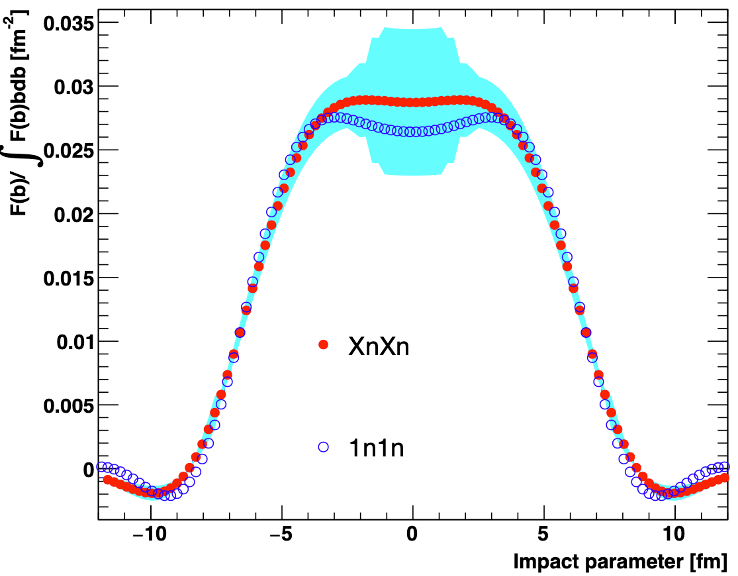}
\caption{(left) $d\sigma/dt$ for coherent photoproduction of dipions, for both $XnXn$ and $1n1n$ nuclear excitations. $1n1n$ corresponds to the case where each nucleus emits a single neutron, usually via Giant Dipole Resonance excitation, while $XnXn$ corresponds to observing any non-zero number of neutrons from each nucleus.
The inset shows a blow-up at very small $|t|$, where destructive interference between the production on the two nuclei \cite{Klein:1999gv} is visible.  At slightly larger $|t|$, the data is reasonably well fit to an exponential, in a limited $|t|$ range.  At higher $|t|$, two diffractive minima are visible.  (right) F(b), calculated using Eq. \ref{eq:fourier} from the data on the left.  The blue band shows the uncertainty as the $|t|$ range for the Fourier transform is varied. Both figures are from Ref. \cite{STAR:2017enh}.
}
\label{fig:fourier}
\end{figure}

The STAR collaboration used this tomographic approach to determine $F(b)$, using 394,000 photoproduced $\pi^+\pi^-$ pairs \cite{STAR:2017enh}.  Data were collected using a trigger that required neutrons in both zero degree calorimeters, signalling breakup of both nuclei, along with two to six hits in a time-of-flight system that surrounded the STAR time projection chamber.   The dipion mass spectrum was well described by a mixture of $\rho^0$, direct $\pi^+\pi^-$  and $\omega\rightarrow \pi^+\pi^-$, from both coherent and incoherent production.  The collaboration measured the incoherent production in the region $0.2 {\rm GeV}^2 < |t| < 0.45 {\rm GeV}^2$ - above the $|t|$ range where coherent production is significant.  There, $d\sigma/dt$ was fit to a dipole form factor (appropriate for a single-nucleon target).  An exponential did not give a good fit to this data.  The dipole form factor was extrapolated to $t=0$ and subtracted from the total $d\sigma/dt$, leaving the coherent component, shown in Fig. \ref{fig:fourier} (left).  This $d\sigma/dt$ was then Fourier transformed, following Eq. \ref{eq:fourier}, to give the transverse distribution shown in Fig. \ref{fig:fourier} (right).  The blue band shows the uncertainty as the maximum $p_T$ in Eq. \ref{eq:fourier} is varied from 0.05 GeV$^2$ to 0.09 GeV$^2$, showing the effects of windowing.   The maximum $p_T$ has a large effect at small $b$, but does not affect the effective size of the target.  The negative wings at $|b|\approx 9$ fermi are likely due to a contribution from the other nucleus, going in the opposite direction, which contributes a negative amplitude at large $b$. 

There are several issues associated with this first extraction of F(b), beyond the caveats listed above.  The effective nuclear radius is considerably larger than the radius of gold.  This is partly because the measured $d\sigma/dt$ includes contributions from the photon $p_T$ and the detector resolution, in addition to the target.   A follow-up study \cite{Klein:2021mgd} attempted to account for these factors, but the fits preferred an unphysically large gold radius.  The photon $p_T$ spectrum is especially problematic, because it depends on the allowed range of impact parameters; requiring $XnXn$ mutual Coulomb excitation selects events with relatively small impact parameters, with $\langle b \rangle \approx 18$ fermi with gold at RHIC \cite{Baltz:2002pp}.  The limited impact parameter range increases the mean $p_T$ 

The ALICE Collaboration has measured $d\sigma/dt$ for coherent $J/\psi$ photoproduction in lead-lead collisions \cite{ALICE:2021tyx}, also finding a $d\sigma/dt$ compatible with a nuclear size larger than the known Woods-Saxon radius, but consistent with models that include nuclear shadowing or saturation (leading to a nucleus that is effectively larger).  For their study, ALICE removed the effect of detector resolution and the photon $p_T$ (which is much smaller than in the STAR setup).  ALICE has also studied the $p_T$ distribution in incoherent $J/\psi$ photoproduction \cite{ALICE:2023gcs}.  Since the presumptive targets are individual nucleons, the spectrum is harder, but the data indicate that sub-nucleon fluctuations also play a role.  

These studies of the transverse distribution can shed light on GPDs.   There is also another direct GPD measurement, using 
$J/\psi$ photoproduction on polarized targets.  The left-right asymmetry is sensitive to polarized GPDs.  The STAR Collaboration has already made a first measurement of this asymmetry, albeit with large errors and a consistency with zero \cite{Schmidke:2016ccw}.   

Looking ahead, theoretical studies have probed how the full Wigner distribution could be measured in UPCs.  Wigner distributions are functions of two conjugate variables, $p_T$ and $b$, so are problematic to observe.   However, we may be able to evade the uncertainty-principle limits, by using final states that involve two different momentum scales \cite{Hatta:2016dxp}.  Photoproduction of dijets is one example.  In the dipole model, the orientation of the dipole fluctuations follows the photon electric field, which itself follows the impact parameter vector.  The dipole-target cross-section depends on the spatial gradients of parton density, since for finite dipole separation, the quark and the antiquark will see slightly different density distributions.  Correlations between the parton $p_T$ and transverse position lead to an azimuthal correlation between the dijet pair $p_T$ and the difference between the two jet $p_T$, which is related to the dijet pair mass.  This angular distribution is
\begin{equation}
\frac{d\sigma}{d\theta} = v_0 [1 + 2v_2 \cos{2\Delta\theta}]
\end{equation}
where $\theta$ is the angle between $\vec{p}_{T1}+\vec{p}_{T2}$ and $\vec{p}_{T1}-\vec{p}_{T2}$. 

\section{Partonic Fluctuations}

\begin{figure}[t]
\includegraphics[width=.48\textwidth]{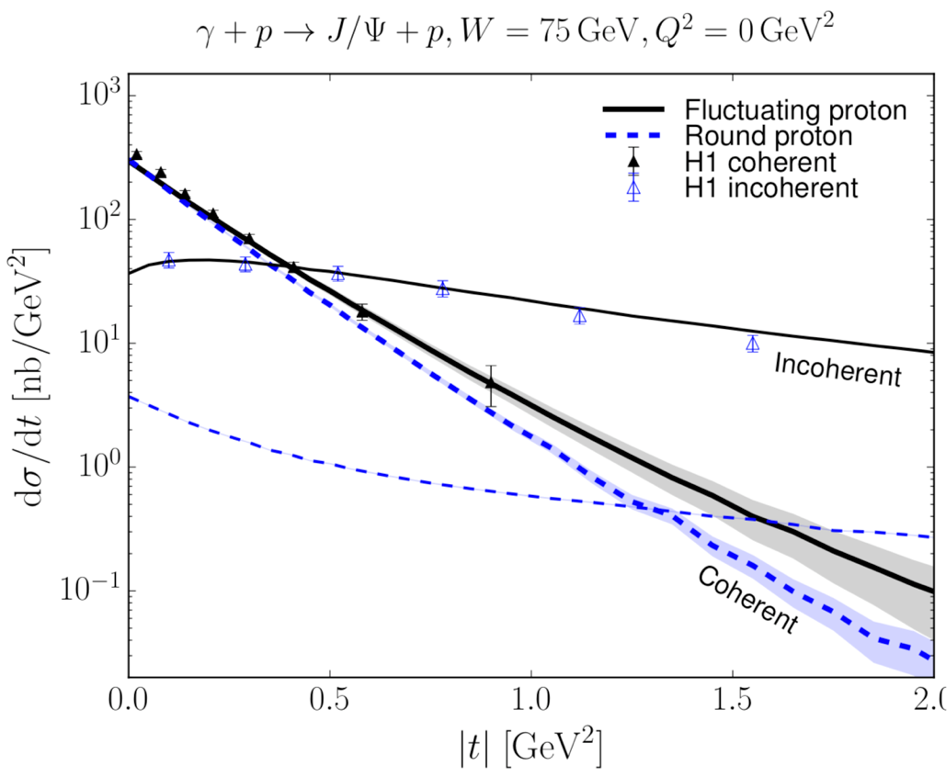}
\includegraphics[width=.48\textwidth]{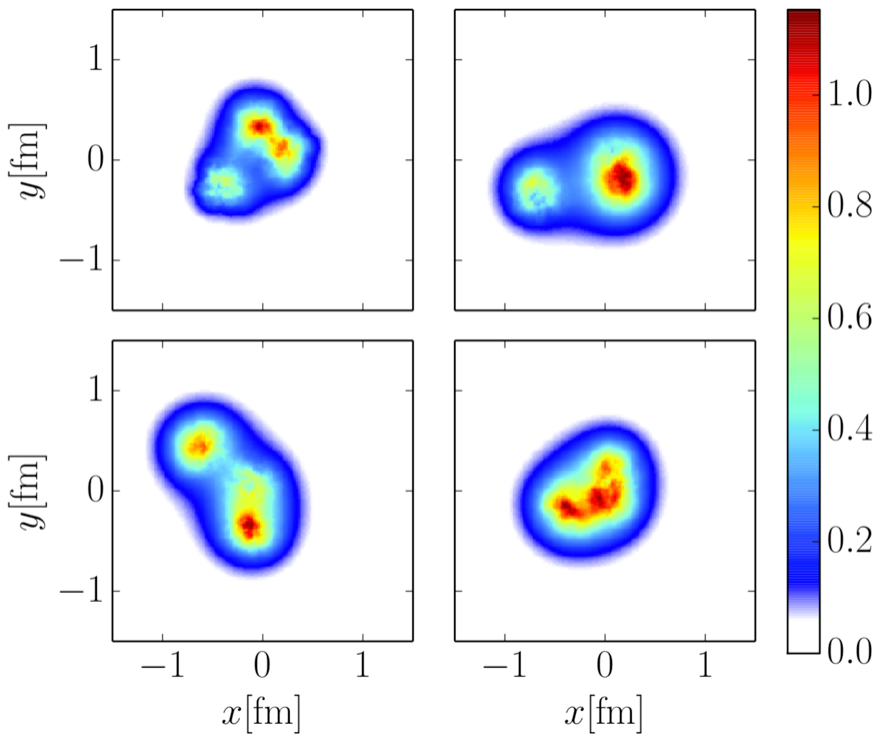}
\caption{(left) HERA data on coherent and incoherent production of $J/\psi$, compared with two models, one with a smooth proton, and the other with a fluctuating proton.  (right) Examples of individual manifestations of the fluctuating proton.  From Ref.  \cite{Mantysaari:2016ykx}.}
\label{fig:fluctuations}
\end{figure}

As noted above, the Good-Walker approach links $d\sigma_{\rm incoherent}/dt$ to fluctuations in the target, including both variations in the spatial distribution of nucleons, and also partonic flucutations, such as an evanescent gluonic hotspots.   Although the relationship between $t$ and distance scales is not as clear as for coherent production, the incoherent $d\sigma/dt$ is well suited to model testing.  Heikki Mantysaari and Bjorn Schenke fitted coherent and incoherent $J/\psi$ photoproduction data from HERA to two models \cite{Mantysaari:2016ykx}.  One model was a smooth (Gaussian) proton, while the other had a fluctuating proton with fluctuating parton configurations.   As Fig. \ref{fig:fluctuations} shows, the fluctuating proton fit the data, while the smooth proton had a much smaller incoherent cross section than the data.  The figure also shows some examples of the fluctuating proton. 

The energy evolution of the incoherent cross-section lends further clues to nuclear evolution \cite{Cepila:2018zky}.  Low photon energies correspond to higher Bjorken$-x$, where fluctuations are expected to be less frequent.  As the energy rises, fluctuations should become more common.  However, as the energy rises further, parts of the target will become fully absorptive, and so will no longer be subject to fluctuations.  At high enough energies, the nucleus is expected to look like a black disk.  At this point, the incoherent cross-section should then be zero.  The energy at which the incoherent cross-section is maximal depends on the mass of the produced vector meson. 

This pattern also applies to nuclei.   Because of the higher parton density in heavy nuclei, the maximum in the incoherent cross-section and the threshold for black disk behavior should be much lower than for proton targets.  

\section{Conclusions} 

Ultra-peripheral collisions are the energy frontier for photon physics, able to probe vector meson production with nearly-real photons up to photon-nucleon center of mass energies of around 2 TeV.  Final states like dijets or open charm are sensitive to the gluon densities in proton or nuclear targets.

Exclusive vector mesons production is slightly more complicated theoretically, requiring two gluons at lowest order.  For proton targets, $\sigma(\gamma p\rightarrow J/\psi p)$ follows a smooth power-law or near-power law increase, with no clear turn-over or other structure.  $J/\psi$ photoproduction on lead targets is suppressed compared to a proton target reference, roughly consistent with the midpoint of nuclear parton distributions, and is indicative of moderate shadowing. 

Because two gluons are exchanged, vector meson photoproduction can be best described using a GPD.   However, the relative energies of the two gluons ($\Delta$) is unknown, so GPDs cannot be directly measured.  However, it is possible to use measurements of $d\sigma/dt$ for coherent photoproduction to study a related quantity, the transverse distribution of gluons in a nucleus.  $J/\psi$ photoproduction data on lead targets is consistent with a moderate difference between the measured transverse gluon distribution and the nucleon distribution in lead, as predicted by models that include shadowing.  

Incoherent photoproduction is sensitive to fluctuations in the nuclear configuration, including fluctuations in parton densities and nucleon positions.  An analysis of coherent and incoherent $J/\psi$ photoproduction data at HERA found that the cross-sections were consistent with a model where the proton underwent fluctuations, whereas a smooth proton model greatly underpredicted the incoherent cross-section.  

Looking ahead, the Electron-Ion Collider will provide very high luminosity $\gamma^*p/\gamma^* A$ collisions which will be studied in detail with an optimized detector - ePIC.  ePIC will offer nearly $4\pi$ coverage, with an extensive forward detector designed to observe the products of nuclear breakup. 

I thank Minjung Kim for useful comments on this manusript.  This work is supported in part by the U.S. Department of Energy, Office of Science, Office of Nuclear Physics, under contract numbers DE-AC02-05CH11231.

\end{document}